\documentclass[conference]{IEEEtran}

\usepackage{mathtools}
\usepackage{SIunits}
\usepackage{ifpdf}
\usepackage{lipsum}
\usepackage{cite}
\usepackage{color}
\usepackage{graphicx}
\usepackage{lipsum}
\usepackage{amsmath}
\usepackage{listings}
\usepackage{xcolor}
\usepackage[linesnumbered,ruled]{algorithm2e}
\usepackage{algpseudocode}
\usepackage{float}
\usepackage{textcomp}
\usepackage{array}
\usepackage{placeins}
\usepackage{stfloats}
\usepackage{url}
\usepackage{placeins}
\usepackage{lipsum}
\usepackage{multirow}
\usepackage{soul}
\usepackage[utf8]{inputenc}
\let\labelindent\relax
\usepackage{enumitem}
\usepackage{algorithmicx}
\graphicspath{{./Figures/}}
\usepackage{glossaries}
\usepackage{diagbox}
\usepackage{tikz}

\title{\LARGE \bf
	Hybrid On/Off Blockchain Approach for Vehicle Data Management, Processing and Visualization Exemplified by the ADAPT Platform}

\author{Aso Validi$^{1}$, \emph{Student Member, IEEE}, Vladislav Kashansky$^{2}$, Jihed Khiari$^{1}$, \emph{Student Member, IEEE}\\  Hamid Hadian$^{2}$, Radu Prodan$^{2}$, Juanjuan Li$^{3}$, Fei-Yue Wang$^{3}$ \\ Cristina Olaverri-Monreal$^{1}$, \emph{Senior Member, IEEE}
	\thanks{$^{1}$Johannes Kepler University Linz, Austria; Chair Sustainable Transport Logistics 4.0. {\tt\small {\{aso.validi, jihed.khiari, cristina.olaverri-monreal\}@jku.at}}}
	\thanks{$^{2}$University of Klagenfurt; Institute of Information Technology.{\tt\small {\{vladislav.kashanskii, hamid.hadian, radu.rodan\}@aau.at}}}%
	\thanks{$^{3}$Institute of Automation Chinese Academy of Sciences China. {\tt\small {\{feiyue.wang, juanjuan.li\}@ia.ac.cn}}}%
}
\begin{document}
	
\IEEEoverridecommandlockouts

\IEEEpubid{
	\begin{minipage}{\textwidth}\ \\\\\\\\\\\\\\\\[12pt] 
		\copyright 2022 IEEE. Personal use of this material is permitted. Permission from IEEE must be obtained for all other uses, in any current or future media, including reprinting/republishing this material for advertising or promotional purposes, creating new collective works, for resale or redistribution to servers or lists, or reuse of any copyrighted component of this work in other works. 
\end{minipage}}
	
\maketitle
\IEEEpubidadjcol	
	
\begin{abstract}
	Hybrid on/off-blockchain vehicle data management approaches have received a lot of attention in recent years. However, there are various technical challenges remained to deal with. In this paper we relied on real-world data from Austria to investigate the effects of connectivity on the transport of personal protective equipment. We proposed a three-step mechanism to process, simulate, and store/visualize aggregated vehicle datasets together with a formal pipeline process workflow model. To this end, we implemented a hybrid blockchain platform based on the hyperledger fabric and gluster file systems. The obtained results demonstrated efficiency and stability for both hyperledger fabric and gluster file systems and ability of the both on/off-blockchain mechanisms to meet the platform quality of service requirements.
\end{abstract}

\section{{Introduction}}
\label{sec:introduction}

Transportation is considered as one of the least visible yet vital elements of the global economy. In today's globalized world, the availability and diversity of products are very heavily depending on the efficiency of transport systems and their capacities. Efficiency and sustainability are the two major challenges that logistics and transport networks are currently dealing with~\cite{hong2007logistics}. The great necessity of a reliable and traceable logistics network has never been more vital than during the current COVID-19 global pandemic. During this period, supply chains are facing several global challenges due to a high demand of sanitary products and in some cases, the closing of borders~\cite{jovanovic2018euro}. According to the World Health Organization (WHO), one of the main strategies to overcome the global shortage of Personal Protective Equipment (PPE) is to coordinate its supply chain management~\cite{Team2020a}.

With the rapid rise of the Internet of Vehicles (IoV) and the automotive sector, using blockchain to manage and store transport and vehicle-related data has attracted academia and industry's attention~\cite{kumarsurvey}. A blockchain-based decentralized framework with a Real-Time Application (RTA) standard provides safe and secure data storing and communication capabilities between vehicles and other entities in the transportation systems~\cite{jabbar2020blockchain, kashansky2021monitoring}.
In this context the \textcolor{black}{ADaptive} and Autonomous data
Performance connectivity and decentralized Transport decision-making network (ADAPT) platform~\cite{kashansky2021adapt, kashansky2021monitoring}, targets to supply, demand, and transport PPE between China and Austria through a transparent, real-time certification verification on equipment, documentation production, and decision-making process at all levels of the multidimensional logistics network.

To this end, we processed real-time data from truck trips in Austria obtained through Otonomo~\cite{otonomo} and replicated two routes in Austria that were relevant to the delivery of PPE using the 3DCoAutoSim simulation platform~\cite{Olaverri-Monreal2018AutomatedAutomation, Capalar2018HypovigilanceWorkload, Olaverri-Monreal2018EffectAvoidance, hussein20183dcoautosim, olaverri2018implementation, olaverri2018connection, artal2019}. We investigated the potential benefits of a  connection between vehicles and stored and visualized the results in our proposed ADAPT hybrid on/off-blockchain platform.
To this purpose, we performed the following three-step approach:

\begin{enumerate}
	\item We first executed a statistical imputation for missing values in machine learning and applied the piecewise cubic Hermite splines (cHs) method to increase the granularity of the dataset by interpolating the missing data points. 
	\item In a second step, we implemented the imputed dataset in the simulation of connected vehicles in platooning mode by relying on the approach in~\cite{validi2021simulation} adopting as well the vehicular communication capabilities of the 3DCoAutoSim simulation platform~\cite{michaeler20173d} to reproduce the platooning connection and analyzed the effects of connected driving in platooning mode on emissions and travel time. 
	\item In a final step, we implemented a hybrid blockchain platform based on the HyperLedger Fabric (HLF) and Gluster File Systems (GlusterFS) to store and visualize the output of the simulation in the ADAPT platform. 
\end{enumerate}

The remaining parts of this paper are organized as follows: the next section presents related work in the field of research. The proposed layered architecture of the hybrid on/off-blockchain model is presented in section~\ref{sec:Multi_ldph}. The data processing and simulation process is described in section~\ref{sec:method_vdps}. Section~\ref{sec:stormodel} presents the on/off-chain storage model. The obtained results are presented in section~\ref{sec:results}. Section~\ref{sec:conclusion} concludes the work and outlines future research.


\section{Related Work}
\label{sec:lr}

There has been a significant increase in research works that evolve around vehicle data analysis and storage in blockchain. In this section we review and categorize the existing approaches as follows:


\subsection{GPS Signal Losses and Solutions}
\label{sec:gps_sls} 

The proliferation of data has enabled many advances in intelligent transportation solutions. Moreover, the quality of the data is essential for the functioning of many of such solutions. On the other hand, the recorded data commonly has issues that can occur along the process of collection, e.g. hardware-related faults, communication/broadcasting faults, or processing mistakes. GPS data can suffer from such issues as well, thus resulting in parts of the trip not having corresponding GPS points. In fact, data imputation or interpolation is a common step when processing GPS data~\cite{shen2014review}. For our use case, it is highly relevant to increase the granularity of the data by interpolating the missing points. This procedure is helpful for the utilization of the GPS data in a simulation downstream. 
A variety of methods have been proposed in literature for GPS data imputation.\cite{rodrigues2018multi} presented a Gaussian process based approach,~\cite{zhang2021imputation} presented a time-series approach named missForest, while~\cite{huang2020integrated} put forth a fuzzy c-means method that is tested with GPS data from taxis. When limited data in terms of density and granularity is available, simpler methods can be considered such as linear interpolation where two consecutive points are joint by a straight line, or splines which are piecewise functions of polynomials allowing to connect two consecutive points by a smooth curve~\cite{de1978practical}. For instance,~\cite{cao2021v} and~\cite{hintzen2010improved} used spline-based methods for GPS interpolation. Similarly, we used the piecewise cubic Hermite spline (cHs) for our task due to the data type and granularity that was available.

\subsection{Storage Methods for Vehicle Data Processing}
\label{sec:storage_m}

\begin{figure}[!t]
	\centering
	\includegraphics[width=0.45\textwidth]{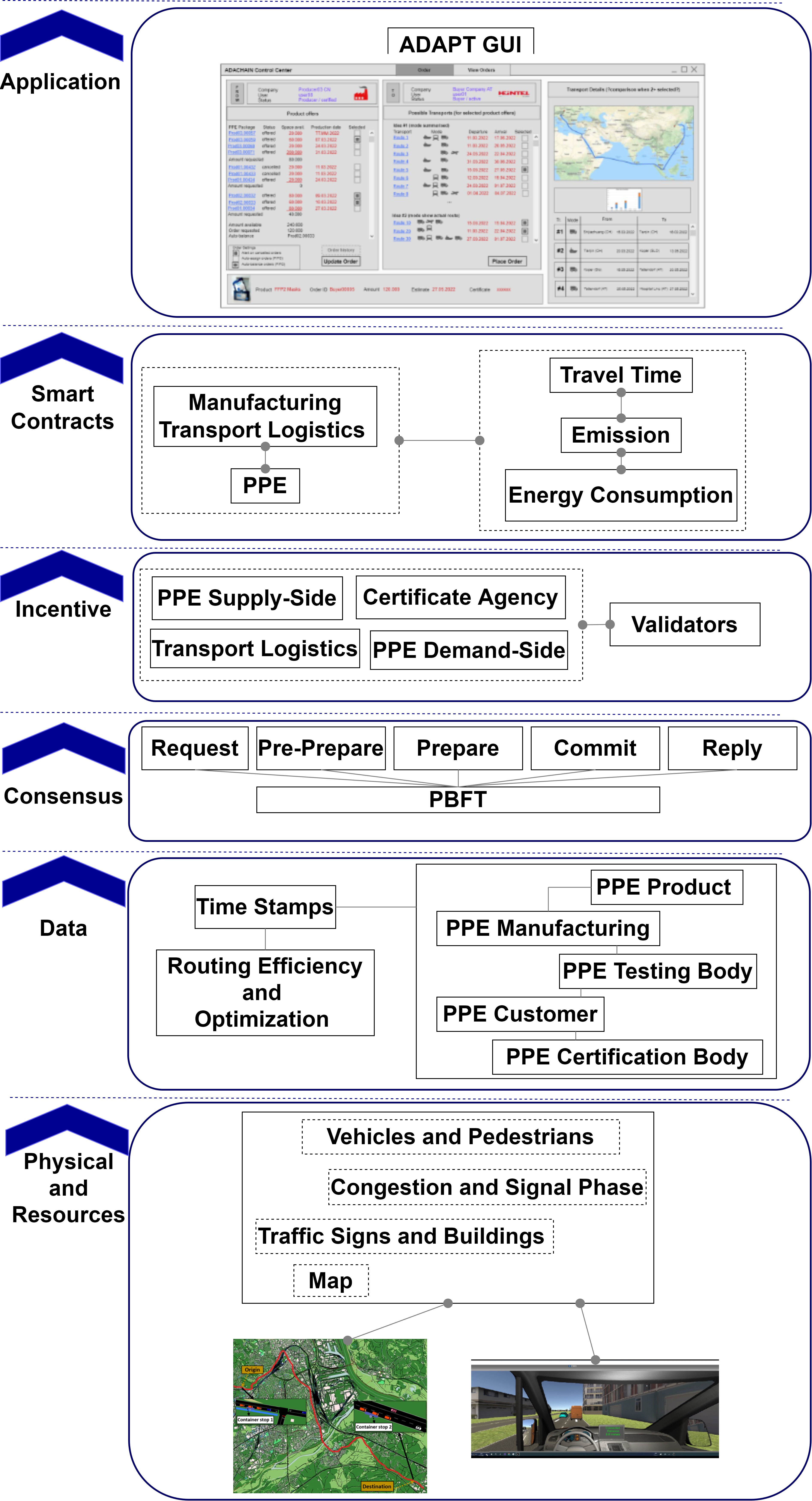}
	\caption{The layered architecture of ADAPT}
	\label{fig:layeredarch}
\end{figure}

Dealing with data volumes that can be processed in combination with a blockchain means to overcome challenges that are either related to the significant data volumes or the huge number of transactions~\cite{ren2018incentive}. There are two fundamentally different approaches for data storage: off-chain, and on-chain. We considered a hybrid combination of the two approaches. In our case, off-chain approach with Distributed File Systems (DFS) is a matter of significant importance, due to the heterogeneity of the various vehicle related data sets. Generally speaking, use case of the distributed file systems in the off-chain scenario differs both quantitatively and qualitatively, and highly depends on the considered use-case. There are several publications on taxonomy, case studies and performance comparisons of distributed file systems~\cite{raicu2011making, depardon2013analysis}. The authors of~\cite{blomer2015experiences} indicated that in data-intensive cases, several different distributed file systems can be used to store different types of files with different access patterns. For example, extended root daemon (XrootD) is optimized for high-performance access to high energy physics datasets~\cite{bird2011computing}. The Hadoop File System (HDFS) is designed for the MapReduce file system infrastructure, and Lustre is optimized for high-performance and highly interoperable applications on supercomputers, which can be combined with Message Passing Interface (MPI) for Input and Output Read (MPI-IO)~\cite{thakur1999implementing, tsujita2016topology}. The authors in~\cite{zhao2013fusionfs, zhao2014fusionfs} studied the behavior of DFS and supporting of data-intensive scientific applications on extreme-scale high-performance computing systems. In this paper we adopted GlusterFS\footnote{Also known as Red Hat Gluster Storage}~\cite{boyer2012glusterfs}. It is highly scalable decentralized file system. 
An important advantage of GlusterFS is that metadata is not stored in a single point of failure, moreover, there is no metadata server (e.g MDS Lustre), which can very often cause the file system to fail. In addition, GlusterFS showed reliable behaviour when servicing data streams of various sizes~\cite{boyer2012glusterfs}.


\section{{Multi-Layer Data Processing Hierarchy}} 
\label{sec:Multi_ldph}

Our proposed ADAPT architecture is a six layered model which is illustrated in Figure~\ref{fig:layeredarch}. It presents the major characteristics and components of the ADAPT hybrid on/off-blockchain platform as follows:

\textbf{\textit{Physical Layer:}} It consists of different sets of physical components within the transport system. 

\textbf{\textit{Data Layer:}} It includes an automated provisioning method for decentralized storage of shared data items with efficient indexing that is based on the GlusterFS system. It incorporates distributed algorithms to balance the blockchain big data storage redundancy overheads, network latency and transaction throughput. Furthermore, it offers secure mechanisms to other services to invoke and fetch stored items.

\textbf{\textit{Consensus Layer:}} The most commonly used consensus in the consortium blockchain is Practical Byzantine Fault Tolerance (PBFT). PBFT is also the consensus algorithm adopted by HLF. There are five steps during the realization of PBFT consensus, which are request, pre-prepare, prepare, commit, and reply.

\textbf{\textit{Incentive Layer:}} It influences the behavior of the system participant. The main objective of the incentive layer is to encourage that all nodes in the hybrid on/off-blockchain system operate in an appropriate way. The PPE-related nodes are mainly PPE supply-side, certificate agencies, transport logistics and PPE demand-side. Furthermore, they rely on validator nodes in the blockchain to help them confirm and record transactions. 

\textbf{\textit {Smart Contracts Layer:}} The key components of this layer are: triggering conditions, response actions, access authorization and communication mode. Smart contracts are a set of self-verifying, self-executing, and self-enforcing state-response rules that are stored and secured in the ADAPT hybrid on/off-blockchain~\cite{yuan2016towards}.

\textbf{\textit{Application Layer:}} It presents the main Graphical User Interface (GUI) operating within our developed ADAPT platform. The GUI presents various types of information for several elements of the logistics' multidimensional network such as PPE documentations, customers and producers. The visualized data consists on the origin-destination and the results from the efficiency analysis of the routes and their related transport modes.


\section{{Vehicle Data Processing and Simulation}}
\label{sec:method_vdps}

\begin{figure}[!t]
	\centering
	\includegraphics[width=0.48\textwidth]{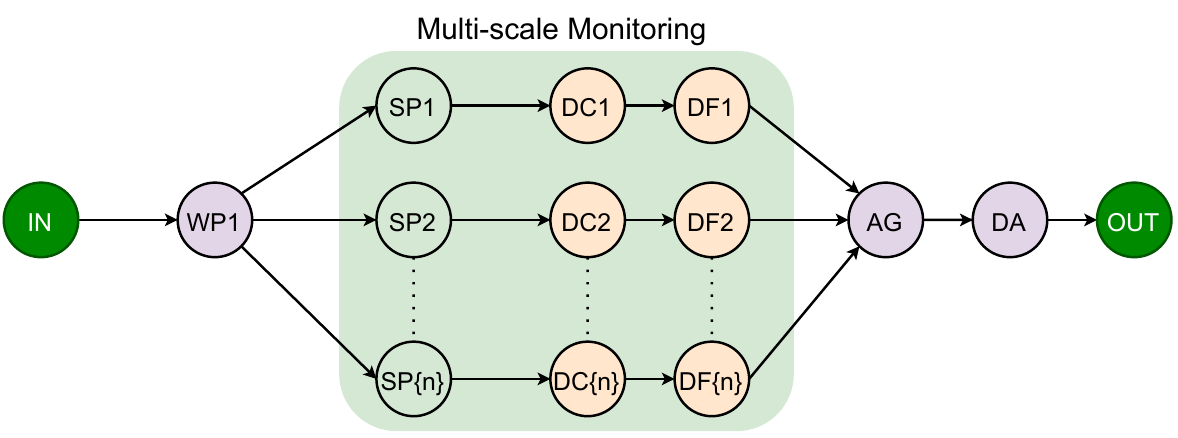}
	\caption{ADAPT multi-scale vehicle data collection workflow~\cite{kashansky2021monitoring}}
	\label{fig:multi_scale_wrkfl}    
\end{figure}

To provide a communication functionality amongst the ADAPT hybrid-blockchain users (transport systems' customers and providers), we visualized various levels of information in the multidimensional logistics network. To this end we first processed, filtered, and filled the missing data in the vehicles' dataset. We then applied the pertinent simulation-based connectivity approaches on the transport system. Finally, we efficiently stored and visualized the data in the proposed ADAPT hybrid-blockchain platform. Figure~\ref{fig:multi_scale_wrkfl} depicts the general structure of the dynamic data collection workflow~\cite{kashansky2021monitoring}. It consists of the following task classes, that are executed over the set of the vehicles $N$:

\begin{description}[font=\normalfont\emph]
	\item[$WP_{1}$] system initialization task, related to the initial configuration of the vehicle data processing workflow;
	\item[$SP_{n}$] activation signal propagation pseudo-task;
	\item[$DC_{n}$] data collection tasks of different duration;
	\item[$DF_{n}$] data filtration tasks (data dependent), which can be optional depending on the transaction instance;
	\item[$AG$] aggregation of results;
	\item[$DA$] data archiving and report preparation;
	\item[$IN$ and $OUT$] source and sink pseudo-tasks of transaction initiator.
\end{description}
Following sections describe semantics of the $DC_{n}$ and $DF_{n}$ task groups.

\subsection{GPS Signal losses Detection and Predicting the Lost Data }
\label{sec:ProcessData}

The acquired data from Otonomo included trips that were undertaken in several countries. We focused in this work on two routes in Austria that were relevant to the delivery of PPE:
\begin{itemize}
	\item $R.VT$:{Vienna Airport to Tattendorf}
	\item $R.TL$:{Tattendorf to Linz}
\end{itemize}
There is a storage facility for PPE in Tattendorf, Linz has a university clinic, and the Vienna Airport is a key distribution point for goods in general and PPE in particular. The Floating Car Data (FCD) dataset contained the following attributes: timestamps, GPS coordinates, speed, and trip ids. 

First, we specified a radius around the origin and destination, and extracted all trips that started and ended within the corresponding radius. For both routes $R.VT$ and $R.TL$, the radius values were $ 3km $ and $ 10km $, respectively. In the case of $R.TL$, the radius was higher due to the lack of availability of data in that area. 

We then selected three trips for each route ($R.VT$ and $R.TL$) based on the density and distribution of points along the route. We labeled the trips for $ R.VT $ as $ T0\_R.VT $, $ T1\_R.VT $ and $ T2\_R.VT $ and for $ R.TL $ as $ T0\_R.TL $, $ T1\_R.TL $ and $ T2\_R.TL $. Each of the trips had the same origin and destination but different middle points. All these trips had varying spatial and temporal granularities, thus resulting in parts of the trips with no GPS coordinates records. To interpolate the missing data, we used piecewise cubic Hermite splines (cHs)~\cite{de1978practical}. By definition, a spline is a function defined piecewise by polynomials. A cubic Hermite spline is a third-degree polynomial specified in Hermite form, i.e. by its values and first derivatives at the end points of the corresponding domain interval. For each two consecutive points in the dataset $i$ and $i+1$, we constructed a curve that was represented by the following expression:
\begin{equation}
	f_i(x)= a_i+b_ix+c_ix^{2}+ d_ix^{3}
\end{equation}
by solving for the four coefficients $a$, $b$, $c$, and $d$. We then pieced together all the curves to obtain a piecewise cubic spline. To do so, cubic Hermite splines ensure $C^{1}$ continuity, meaning that they match slopes at the join points, by matching the first derivatives. For the implementation, we used the \textsc{Python}'s \textsc{SciPy} package, which includes various interpolation functions. After converting the data to the GPS exchange format (GPX), we used the cHs method for interpolation in the longitudinal and latitudinal direction in relation to time and speed with a spatial resolution of $1  meter$. The method generated GPS data for the missing parts of the original trips. The number of interpolated points compared to the original points was about a thousandfold (e.g. trips from Route $R.VT$ originally have $300$ points, cHs generated $300000$ interpolated points). Furthermore, it is fast and computationally inexpensive. On the other hand, this method does not consider information about the underlying road network and can therefore be subject to errors related to the taken road. However, for the purpose of our application, it provided realistic and consistently distributed GPS points that could subsequently be used for simulation.

\subsection{Platooning Implementation }
\label{sec:platoon}

After the previous process we replicated the imputed vehicle data to generate a realistic transport network simulation and analyze the effect of platooning on travel time and emissions. For this purpose, we compared the simulation outcome from three trucks under the following two scenarios:

\begin{enumerate}[label=(\Roman*)]
	\item Connected in platooning mode, the vehicles in this scenario are labeled as: Semi-Autonomous.Leader.Truck.1 ($SAL.{Tr_{1}}$), Autonomous.Follower.Truck.2 ($AF.{Tr_{2}}$) and Autonomous.Follower.Truck.3 ($AF.{Tr_{3}}$).
	\item Not connected (conventional driving, not platooning), the vehicles in this scenario are labeled as: Truck.1 ($Tr_{1}$), Truck.2 ($Tr_{2}$) and  Truck.3 ($Tr_{3}$).
\end{enumerate}

To this end, we implemented consistent platooning~\cite{validi2021simulation} and calculated the travel time and total emissions according to the following steps.

\subsubsection{Traffic Network}
\label{sec:trafficnetwork}

To generate Austria's traffic network we utilized the method explained in~\cite{validi2021analysis, validi2021simulation}. Following the adopted method, we imported $.osm$ data from Open Street Map (OSM) to the Simulation of Urban Mobility (SUMO)~\cite{behrisch2011sumo} and applied a variety of SUMO applications~\cite{Netconvert, Polyconvert, NeteditDocumentation} to generate and visualize the traffic network. 
Additionally, to create routes and traffic demand we adopted two different methods and generated two sets of route files. Each method and the corresponding obtained route files were dedicated to each specific dataset. 

We applied $tracemapper.py$~\cite{RoutesSumo} to our first imputed GPS dataset. Here the input file ($latlong.xml$) contained the latitude and longitude coordinates of the vehicles. The output file ($latlong.rou.xml$) was a SUMO route file based on the geographical coordinates. This method mapped a list of geo coordinates to a consecutive list of edges in a given network. 
\begin{algorithm}[!t]
	\small
	\caption{\small Generating the route files and vehicles~\cite{validi2021simulation}} 
	\DontPrintSemicolon
	\label{algo:routefiles}
	\SetAlgoLined
	\SetKwData{Left}{left}\SetKwData{This}{this}\SetKwData{Up}{up}
	\SetKwInOut{Input}{input}\SetKwInOut{Output}{output}\SetKwInOut{Define}{define}
	\Input{Network, $r$;}
	\Output{XYRoute file, $rou$;}
	\Define{Number of vehicles $i$, Number of time steps $n$, Vehicle types; platooningvType, truckvType, truckClass;}
	\SetKwFunction{FMain}{RoutVehGenerate}
	\SetKwProg{Fn}{Function}{:}{End~Function}
	\Fn{\FMain{}}
	{ {random.seed(s)} \\
		{N $\leftarrow$ n} \\
		\Output{$.rou.xml$ ($vType$, edges, $veh_{id}$)} \; 
		vehNr $\leftarrow$ 0 \;
		\For{i in range(N)}{
			\If{random.uniform(0,1) $<$ each of the three trucks:}{
				write output'$<$ generating the vehicle types with different attributes$/>$' format ($i,vehNr$), file = routes\; 
				vehNr $+= 1$}
		}
	}          	
\end{algorithm}

\begin{figure}[!t]
	\centering
	\includegraphics[width=0.47\textwidth]{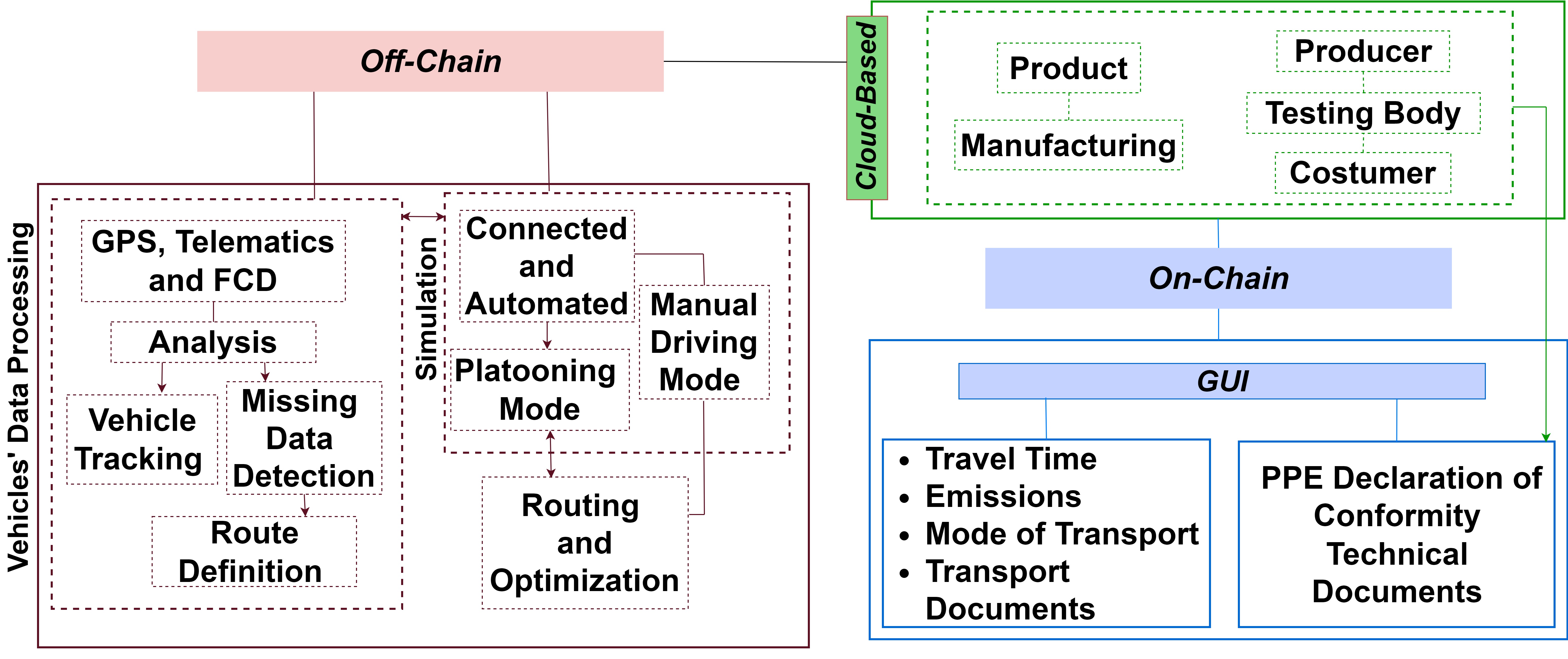}
	\caption{Hybrid on/off-chain storage architecture}
	\label{fig:storing_m}
\end{figure}

The process for generating the second set of route files and the traffic demand using Traffic Control Interface (TraCI)~\cite{traciSUMO} is described in Algorithm~\ref{algo:routefiles}~\cite{validi2021simulation}. The second set of route files ($XY.rou.xml$) was generated based on the $XY coordinates$ of vehicles in the simulation network ($.net.xml$). A calibration of the traffic flow was then performed.

\subsubsection{Vehicle Connection}
\label{sec:vehicleconnection}

By accessing related nodes in Objective Modular Network Testbed in C++ (OMNeT++) and changing the behavior of the vehicles, we reproduced the simulation of connected vehicles in platooning mode. Furthermore, as the simulation necessitates a large computing capacity, to make it functioning as smoothly as possible, we created a small number of SUMO vehicles that were not identified as OMNeT++ nodes and lacked the communication capabilities. To investigate the potential and beneficial effects of driving in platooning mode, we conducted a set of comparative analyses between connected and not connected trucks. To this end, we analyzed travel time and total emissions obtained from the simulations. To study the total emissions we relied on the ``$ HBEFA3/HDV\_D\_(EU5) $'' emission class that includes the weight and length of the trucks ($ 26\_40t, EURO~5 $) used~\cite{HBEFA}. The analyzed data (for each trip) were then stored and visualized in the ADAPT hybrid-blockchain platform as described in the next section.

\begin{figure}[!t]
	\centering
	\includegraphics[scale=0.35]{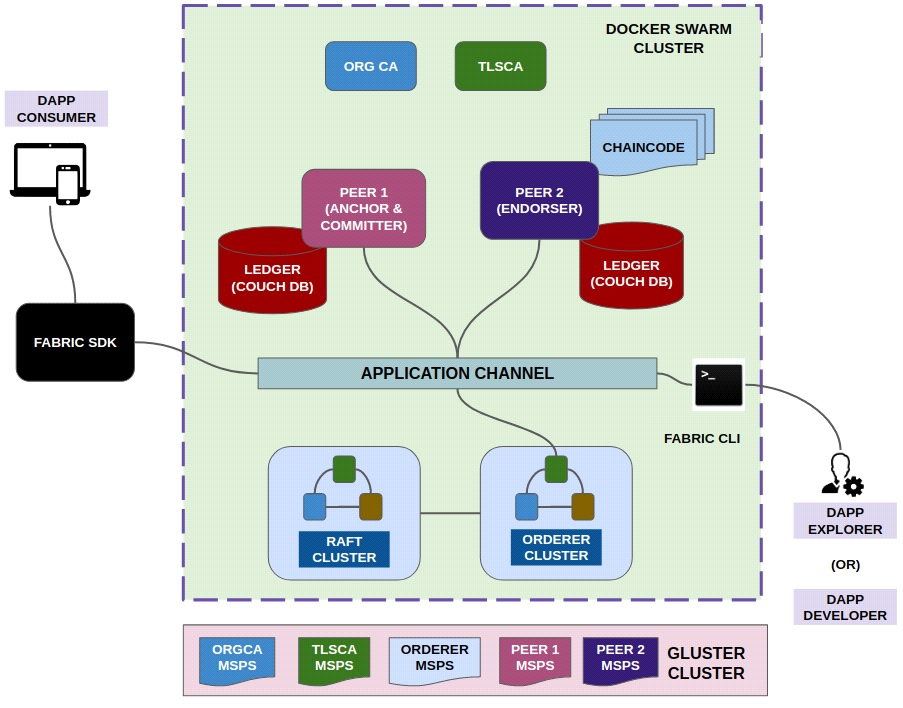}
	\caption{ADACHAIN service configuration architecture for vehicle data management~\cite{kashansky2021adapt}}
	\label{fig:adachain-bc-arch-1}    
\end{figure}

\section{Hybrid On/Off-Chain Storage Model}
\label{sec:stormodel}

\subsection{Model Implementation}

\begin{table*}[!t]
	\centering
	\caption{Efficiency in studied scenarios and routes}
	\renewcommand{\arraystretch}{1.51}
	\label{table:summarykeymetrics}
	\resizebox{\linewidth}{!}{%
		\begin{tabular}{|l|l|c|c|c|c|c|c|l|c|l|c|c|c|l|c|l|c|l|c|} 
			\cline{3-20}
			\multicolumn{1}{l}{} &  & \multicolumn{6}{c|}{\begin{tabular}[c]{@{}c@{}}\textbf{Travel Time}\\$[s]$\end{tabular}} & \multicolumn{12}{c|}{\begin{tabular}[c]{@{}c@{}}\textbf{Total Emissions}\\\textbf{(cumulated)}\\$[mg/s]$\end{tabular}} \\ 
			\cline{2-20}
			\multicolumn{1}{c|}{} & \multicolumn{1}{c|}{\textbf{Route}} & \multicolumn{3}{c|}{\textbf{R.VT}} & \multicolumn{3}{c|}{\textbf{R.TL}} & \multicolumn{6}{c|}{\textbf{\textbf{R.VT}}} & \multicolumn{6}{c|}{\textbf{\textbf{R.TL}}} \\ 
			\cline{2-20}
			\multicolumn{1}{c|}{} & \multicolumn{1}{c|}{\textbf{Trip}} & \begin{tabular}[c]{@{}c@{}}T0\_R.VT \\44.3$km$\end{tabular} & \begin{tabular}[c]{@{}c@{}}T1\_R.VT \\43.4$km$\end{tabular} & \begin{tabular}[c]{@{}c@{}}T2\_R.VT \\43.1$km$\end{tabular} & \begin{tabular}[c]{@{}c@{}}\textbf{T0\_R.TL }\\\textbf{212$km$}\end{tabular} & \begin{tabular}[c]{@{}c@{}}T1\_R.TL \\209$km$\end{tabular} & \begin{tabular}[c]{@{}c@{}}T2\_R.TL \\207$km$\end{tabular} & \multicolumn{2}{c|}{\begin{tabular}[c]{@{}c@{}}T0\_R.VT \\44.3$km$\end{tabular}} & \multicolumn{2}{c|}{\begin{tabular}[c]{@{}c@{}}T1\_R.VT \\43.4$km$\end{tabular}} & \multicolumn{2}{c|}{\begin{tabular}[c]{@{}c@{}}T2\_R.VT \\43.1$km$\end{tabular}} & \multicolumn{2}{c|}{\begin{tabular}[c]{@{}c@{}}T0\_R.TL \\212$km$\end{tabular}} & \multicolumn{2}{c|}{\begin{tabular}[c]{@{}c@{}}T1\_R.TL \\209$km$\end{tabular}} & \multicolumn{2}{c|}{\begin{tabular}[c]{@{}c@{}}T2\_R.TL \\207$km$\end{tabular}} \\ 
			\hline
			\multirow{3}{*}{\begin{tabular}[c]{@{}l@{}}\textbf{Connected}\\(in platooning mode)\end{tabular}} & \textbf{$SAL.{Tr_{1}}$} & \multirow{3}{*}{\textbf{2534}} & \multirow{3}{*}{\textbf{2532}} & \multirow{3}{*}{\textbf{2530}} & \multirow{3}{*}{\textbf{7898}} & \multirow{3}{*}{\textbf{7697}} & \multirow{3}{*}{\textbf{7624}} & \multicolumn{1}{c|}{\textbf{\textbf{3720.76}}} & \multirow{3}{*}{\begin{tabular}[c]{@{}c@{}}\textbf{\textbf{}}\\\textbf{\textbf{SUM =}}\\\textbf{\textbf{10956.31}}\\\textbf{}\end{tabular}} & \multicolumn{1}{c|}{\textbf{\textbf{\textbf{\textbf{3719.56}}}}} & \multirow{3}{*}{\begin{tabular}[c]{@{}c@{}}\textbf{SUM =}\\\textbf{10947.66}\end{tabular}} & \textbf{\textbf{\textbf{\textbf{\textbf{\textbf{\textbf{\textbf{3717.24}}}}}}}} & \multirow{3}{*}{\begin{tabular}[c]{@{}c@{}}\textbf{\textbf{SUM =}}\\\textbf{\textbf{10939.01}}\end{tabular}} & \textbf{11482.54} & \multirow{3}{*}{\begin{tabular}[c]{@{}c@{}}\textbf{\textbf{\textbf{\textbf{SUM =}}}}\\\textbf{\textbf{\textbf{\textbf{34148.74}}}}\end{tabular}} & \textbf{\textbf{11371.81}} & \multirow{3}{*}{\begin{tabular}[c]{@{}c@{}}\textbf{\textbf{\textbf{\textbf{SUM=}}}}\\\textbf{\textbf{\textbf{\textbf{\textbf{\textbf{\textbf{\textbf{33279.68}}}}}}}}\end{tabular}} & \textbf{\textbf{\textbf{\textbf{11253.09}}}} & \multirow{3}{*}{\begin{tabular}[c]{@{}c@{}}\textbf{\textbf{\textbf{\textbf{SUM =}}}}\\\textbf{\textbf{\textbf{\textbf{32964.04}}}}\end{tabular}} \\ 
			\cline{2-2}\cline{9-9}\cline{11-11}\cline{13-13}\cline{15-15}\cline{17-17}\cline{19-19}
			& \textbf{$AF.{Tr_{2}}$ } &  &  &  &  &  &  & \textbf{\textbf{3652.12}} &  & \textbf{\textbf{\textbf{\textbf{3649.36}}}} &  & \textbf{\textbf{\textbf{\textbf{\textbf{\textbf{\textbf{\textbf{3648.77}}}}}}}} &  & \textbf{\textbf{11377.32}} &  & \textbf{\textbf{\textbf{\textbf{11120.84}}}} &  & \textbf{\textbf{\textbf{\textbf{\textbf{\textbf{\textbf{\textbf{11176.45}}}}}}}} &  \\ 
			\cline{2-2}\cline{9-9}\cline{11-11}\cline{13-13}\cline{15-15}\cline{17-17}\cline{19-19}
			& \textbf{$AF.{Tr_{3}}$ } &  &  &  &  &  &  & \textbf{\textbf{\textbf{\textbf{3583.43}}}} &  & \textbf{3578.74} &  & \textbf{3573} &  & \textbf{11288.88} &  & \textbf{10787.03} &  & \textbf{10534.52} &  \\ 
			\hline
			\multirow{3}{*}{\begin{tabular}[c]{@{}l@{}}\textbf{Not connected}\\\textbf{(}conventional driving, \\not platooning\textbf{)}\end{tabular}} & \textbf{$Tr_{1}$} & \multirow{3}{*}{\textbf{3235}} & \multirow{3}{*}{\textbf{3233}} & \multirow{3}{*}{\textbf{3230}} & \multirow{3}{*}{\textbf{10081}} & \multirow{3}{*}{\textbf{9827}} & \multirow{3}{*}{\textbf{9725}} & \textbf{4638.78} & \multirow{3}{*}{\begin{tabular}[c]{@{}c@{}}\textbf{SUM =}\\\textbf{13278.38}\end{tabular}} & \textbf{\textbf{4636.92}} & \multirow{3}{*}{\begin{tabular}[c]{@{}c@{}}\textbf{\textbf{SUM =}}\\\textbf{\textbf{13270.47}}\end{tabular}} & \multicolumn{1}{l|}{\textbf{\textbf{\textbf{\textbf{4634.63}}}}} & \multirow{3}{*}{\begin{tabular}[c]{@{}c@{}}\textbf{\textbf{SUM =}}\\\textbf{\textbf{13258.16}}\end{tabular}} & \textbf{\textbf{13823.21}} & \multirow{3}{*}{\begin{tabular}[c]{@{}c@{}}\textbf{\textbf{\textbf{\textbf{SUM =}}}}\\\textbf{\textbf{\textbf{\textbf{}}41386.65\textbf{\textbf{}}}}\end{tabular}} & \textbf{\textbf{\textbf{\textbf{13803.21}}}} & \multirow{3}{*}{\begin{tabular}[c]{@{}c@{}}\textbf{\textbf{\textbf{\textbf{SUM =}}}}\\\textbf{\textbf{\textbf{\textbf{40343.91}}}}\end{tabular}} & \textbf{\textbf{\textbf{\textbf{\textbf{\textbf{\textbf{\textbf{13685.53}}}}}}}} & \multirow{3}{*}{\begin{tabular}[c]{@{}c@{}}\textbf{\textbf{\textbf{\textbf{SUM =}}}}\\\textbf{\textbf{\textbf{\textbf{39925.16}}}}\end{tabular}} \\ 
			\cline{2-2}\cline{9-9}\cline{11-11}\cline{13-13}\cline{15-15}\cline{17-17}\cline{19-19}
			& \textbf{$Tr_{2}$} &  &  &  &  &  &  & \textbf{\textbf{\textbf{\textbf{4526.23}}}} &  & \textbf{\textbf{\textbf{\textbf{\textbf{\textbf{\textbf{\textbf{4524.77}}}}}}}} &  & \multicolumn{1}{l|}{\textbf{\textbf{\textbf{\textbf{\textbf{\textbf{\textbf{\textbf{\textbf{\textbf{\textbf{\textbf{\textbf{\textbf{\textbf{\textbf{4522.87}}}}}}}}}}}}}}}}} &  & \textbf{\textbf{13815.55}} &  & \textbf{\textbf{\textbf{\textbf{13800.88}}}} &  & \textbf{\textbf{\textbf{\textbf{\textbf{\textbf{\textbf{\textbf{13329.16}}}}}}}} &  \\ 
			\cline{2-2}\cline{9-9}\cline{11-11}\cline{13-13}\cline{15-15}\cline{17-17}\cline{19-19}
			& \textbf{$Tr_{3}$} &  &  &  &  &  &  & \textbf{4213.37} &  & \textbf{4108.78} &  & \multicolumn{1}{l|}{\textbf{4100.75}} &  & \textbf{\textbf{13747.89}} &  & \textbf{12739.82} &  & \textbf{12910.47} &  \\
			\hline
		\end{tabular}
	}
\end{table*}

\begin{figure}[!t]
	\centering
	\includegraphics[width=0.45\textwidth]{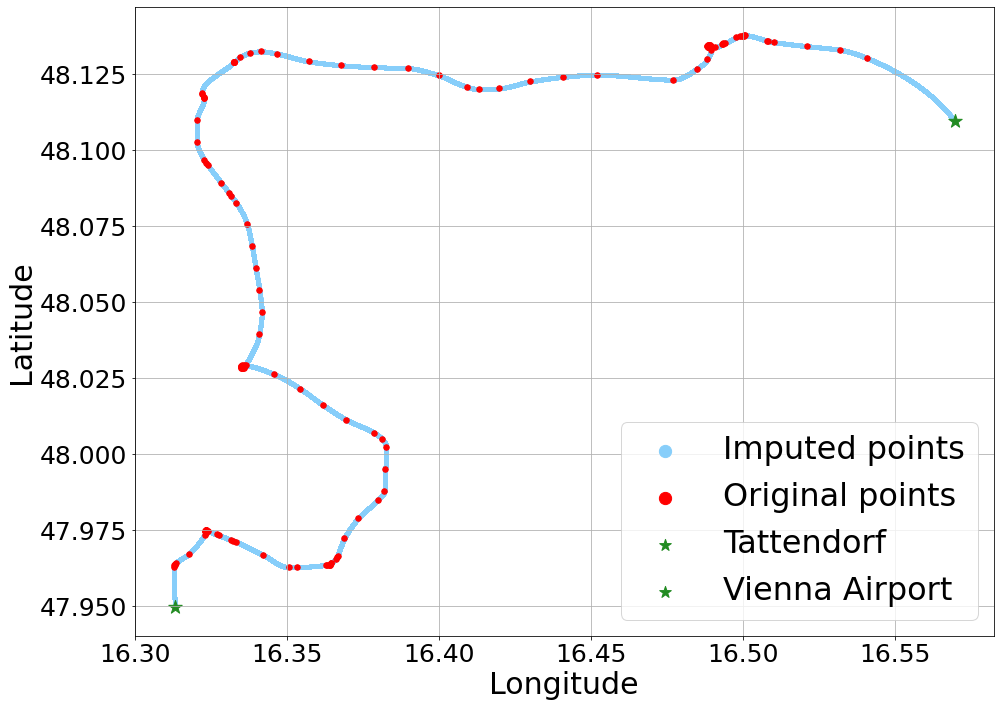}
	\caption{Imputation results for $T0\_R.VT$}
	\label{fig:imp_r1}
\end{figure}

\begin{figure*}[!t]
	\centering
	\vspace*{-70mm}
	\includegraphics[width=\textwidth]{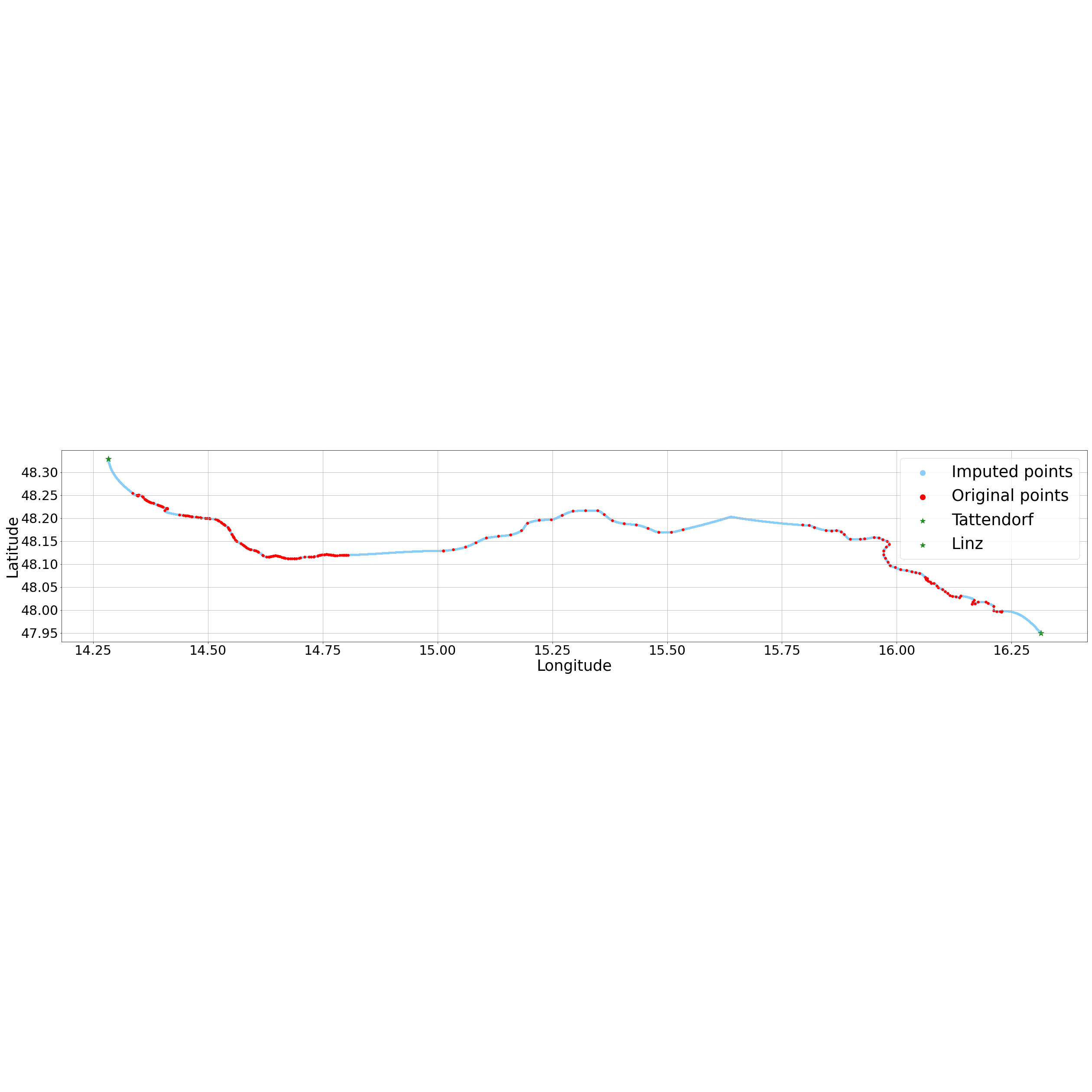}
	\vspace*{-70mm}
	\caption{Imputation results for $T0\_R.TL$ }
	\label{fig:imp_r2}
\end{figure*}

Figure~\ref{fig:storing_m} illustrates our proposed ADAPT on/off-chain storing model. The on-chain functionality is provided by HLF fabric and corresponding smart contracts. Off-chain functionality is driven by the distributed file system and REpresentational State Transfer (REST) API, that we use for storing of structured and unstructured data. The data used in smart contracts are cloud-based distributed. In this case on-chain PPE data storage is not applicable for two main reasons. First, part of the data cannot be disclosed for privacy concerns. Second, a few data owners (e.g. companies, government organizations) are restricted by particular policies to store their information on blockchain. As a consequence, a hybrid on/off-chain solution seemed here to be more appropriate. This prescribes that some parts of the data are stored off-chain, and some parts are stored on-chain. The off-chain data is stored in a Gluster distributed file system, meanwhile corresponding hash values and index information is recorded into the blockchain to ensure authenticity of the data.

We adopted ADACHAIN (Fig.~\ref{fig:adachain-bc-arch-1}) as it uses GlusterFS~\cite{boyer2012glusterfs} for the persistent storage and data management of common services in addition to the transaction ledger databases. GlusterFS unifies all storage nodes located in various servers as a single file system. This network-attached storage offers high-performance storage support and helps address  high availability, resilience, auto repair, and automatic recovery of the decentralized logistic services during system or infrastructure failures. From the operating system point of view, GlusterFS mounts the file system via Filesystem in Userspace (FUSE) and supports Portable Operating System Interface (POSIX). This facilitates developing standard applications. 

\subsection{Verification Process}
To verify our proposed off-chain storage model, we conducted preliminary experiments with approximately $14 GB$ of raw data in combined exchange regime with some \textcolor{black}{Input and Output zone} (IOzone)~\cite{norcott2003iozone} benchmarks. 
The experimental environment for data management strategies was provided by cloud services of the distributed systems group at the Alpen-Adria university of Klagenfurt.
We first conducted a simulation experiment of incremental update based on the simulation of data and compared it with the traditional update solution. We then analyzed the additional time required to implement the incremental update. Finally, we evaluated the performance of the off-chain storage system and analyzed the impact of different operations and data exchange patterns in the context of overall system's performance.

\section{{Results from the Analysis and Verification Process}}
\label{sec:results}

Figures~\ref{fig:imp_r1} and~\ref{fig:imp_r2} depict the imputation results we obtained by using the cHs method for two sample trips $T0\_R.VT$ and $T0\_R.TL$ from the considered routes $R.VT$ and $R.TL$. We noted the original points in red and the imputed ones in blue. The origins and destinations are shown in green. The imputed points connect the original ones throughout the routes with a spatial granularity of 1 meter, thus resulting in more detailed and consistent GPS data. A summary of all the obtained values for travel time and total emissions in each of the three trips of the two routes are presented in Table~\ref{table:summarykeymetrics}. This information was visualized in the hybrid on/off-blockchain ADAPT platform through the GUI, providing thus the different sectors of the multidimensional logistics network with valuable information and different options regarding efficient and sustainable transportation.

The overall results demonstrated significant system's stability for both on/off-chain HLF and GlusterFS mechanisms to meet the required quality of Service (QoS). Table~\ref{table:gfs_test_metrics} presents parts of the GlusterFS volume profile per-brick I/O information for each File Operation (FOP) of a volume. The per brick information provides the possibility to identify the bottlenecks in the storage system.

\begin{figure}[!t]
\centering
\includegraphics[width=0.5\textwidth]{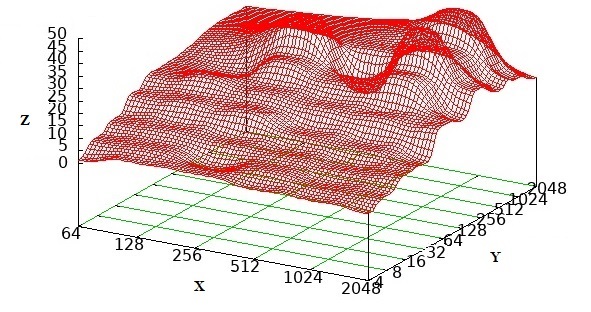}
\caption{Preliminary performance data, collected with the  IOzone~\cite{norcott2003iozone} system for~\textit{WRITE} off-chain operations with GlusterFS distributed filesystem as part of the ADACHAIN framework; Z-axis - write speed in MBytes/s, X-axis - file size in KBytes/s, Y-axis - record size in KBytes/s}
\label{fig:gfs_w_op_perf_iozone}    
\end{figure}

Figure~\ref{fig:gfs_w_op_perf_iozone} illustrates the  non-linear performance effects of data processing patterns in the context of the \textit{WRITE} operation performance obtained from the IOzone~\cite{norcott2003iozone} benchmark suite in combined vehicle data processing regime. It presents the initial results as the relation between the particular file sizes and the variation of the record sizes $y \in [64, 2048]$ and the aggregators $x \in [4, 2048]$, corresponding with the Z-axis to the performance in KBytes/sec. In general, the storage system scales properly, and demonstrates high reliability and robustness.

\begin{table}[!t]
\centering
\caption{GlusterFS volume profile per-brick I/O information for each File Operation Platform (FOP) of a volume. The per brick information helps in identifying bottlenecks in the storage system. Result of the gluster volume profile info }
\label{table:gfs_test_metrics}
\renewcommand{\arraystretch}{1.3}
\resizebox{\linewidth}{!}{%
	\begin{tabular}{llllll} 
		\hline
		\begin{tabular}[c]{@{}l@{}}\\\%-latency\end{tabular} & AVG (\micro\second) & MIN (\micro\second) & MAX (\micro\second) & \# of calls & Operation \\ 
		\hline
		4.82 & 1132.28 & 21.00 & 800970.00 & 4575 & WRITE \\ 
		\hline
		5.70 & 156.47 & 9.00 & 665085.00 & 39163 & READDIRP \\ 
		\hline
		11.35 & 315.02 & 9.00 & 1433947.00 & 38698 & LOOKUP \\ 
		\hline
		11.88 & 1729.34 & 21.00 & 2569638.00 & 7382 & FXATTROP \\ 
		\hline
		47.35 & 104235.02 & 2485.00 & 7789367.00 & 488 & FSYNC \\
		\hline
	\end{tabular}
}
\end{table}

\section{{Conclusion and Future Work}}
\label{sec:conclusion}
In this paper we presented a three-step mechanism
relying on real-world data from Austria to process, simulate, and visualize aggregated vehicle data in a hybrid on/off-blockchain platform. To this end we relied on the 3DCoAutoSim Simulation Platform and hybrid blockchain platform based on the hyperledger fabric and gluster file systems. The obtained results demonstrated  efficiency and stability for both hyperledger fabric and gluster file systems and ability of the both on/off-blockchain mechanisms to meet the platform quality of service requirements.
In future work we aim at extending our research by developing bilateral interaction between SUMO and the blockchain to investigate different Intelligent Transportation Systems (ITS) applications. We will additionally investigate particular I/O limitation policies and asymmetry scenarios between storage peers and corresponding clients.


\section*{Acknowledgment}
This work was supported by ADaptive and Autonomous data Performance connectivity and decentralized Transport decision-making network (ADAPT) project funded by ICT of the Future: 6th Call for Cooperative R\&D Projects between Austria, the austrian research promotion agency (FFG), and China, Chinese Academy of Sciences (CAS).


\bibliographystyle{IEEEtran}
\bibliography{references}

\end{document}